\documentclass[preprint,showkeys,amssymb,prd,aps,onecolumn,amsfonts,eqsecnum,nofootinbib]{revtex4}

\usepackage{amsfonts}
\usepackage{amssymb}
\usepackage{graphicx}
\usepackage{amsmath}                  
\usepackage{color}
\usepackage{graphics,epsfig}

\begin{document}

\newcommand {\dttaa} {{\bar\theta}^{\dot \alpha}}
\newcommand {\dttab} {{\bar\theta}^{\dot \beta}}
\newcommand {\dttag} {{\bar\theta}^{\dot \gamma}}
\newcommand {\ttaa} {\theta^\alpha}
\newcommand {\ttab} {\theta^\beta}
\newcommand {\ttag} {\theta^\gamma}
\newcommand {\tta} {\theta}
\newcommand {\btta} {\bar \theta}

\newcommand {\smaa} {\sigma^\mu_{\alpha \dot\alpha}}
\newcommand {\smbb} {\sigma^\mu_{\beta \dot\beta}}
\newcommand {\smgg} {\sigma^\mu_{\gamma \dot\gamma}}
\newcommand {\snaa} {\sigma^\nu_{\alpha \dot\alpha}}
\newcommand {\snbb} {\sigma^\nu_{\beta \dot\beta}}
\newcommand {\sngg} {\sigma^\nu_{\gamma \dot\gamma}}
\newcommand {\cab}  {C^{\alpha\beta}}
\newcommand {\cgd}  {C^{\gamma\delta}}
\newcommand {\bcab} {\bar{C}^{\dot\alpha\dot\beta}}
\newcommand {\bcgd} {\bar{C}^{\dot\gamma\dot\delta}}
\newcommand {\al} {\alpha}
\newcommand {\bt} {\beta}
\newcommand {\vt} {\bigg\vert}
\newcommand {\be} {\begin{equation}}
\newcommand {\bear} {\begin{eqnarray}}
\newcommand {\ee} {\end{equation}}
\newcommand {\eear} {\end{eqnarray}}
\newcommand {\besp} {\begin{equation}\begin{split}}
\newcommand {\eesp} {\end{split}\end{equation}}

\preprint{WM-04-120}
\title{\vspace*{0.5in} A Field Theoretical Model in Noncommutative Minkowski Superspace~\footnote{Presented at the APS's 2004 Meeting of the Division of Particles and Fields, University of California, Riverside, 26-31 August 2004. This talk will be published in the proceedings of DPF 2004 as a supplement in International Journal of Modern Physics A.}
\vskip 0.1in}
\author{Vahagn Nazaryan}\email[]{vrnaza@wm.edu}
\author{Carl E. Carlson}\email[]{carlson@physics.wm.edu}
\affiliation{Particle Theory Group, Department of Physics,
College of William and Mary, Williamsburg, VA 23187-8795}
\date{November 2004}
\begin{abstract}
In this talk we present a field theoretical model constructed in Minkowski $\mathcal{N}=1$ superspace with a deformed supercoordinate algebra. Our study is motivated in part by recent results from super-string theory, which show that in a particular scenario in Euclidean superspace the spinor coordinates $\tta$ do not anticommute. Field theoretical consequences of this deformation were studied in a number of articles. We present a way to extend the discussion to Minkowski space, by assuming non-vanishing anticommutators for both $\tta$, and $\btta$. We give a consistent supercoordinate algebra, and a star product that is real and preserves the (anti)chirality of a product of (anti)chiral superfields. We also give the Wess-Zumino Lagrangian $\mathcal{L}_{WZ}$ that gains Lorentz-invariant corrections due to non(anti)commutativity within our model. The Lagrangian in Minkowski superspace is also always manifestly Hermitian.

\end{abstract}
\maketitle



\section{Introduction}

There is a considerable amount of discussion in the literature during
recent years about a possibly richer, nontrivial structure of
space-time. In many of the scenarios the space-time coordinates $x^\mu$
become noncommutative. The studies of theories with noncommutative space-time gained motivation in recent years predominantly form the observation that string theories in a background field can be solved exactly and give coordinate operators which do not commute~\cite{ardalan,SW}. The parameter $\Theta^{\mu\nu}$ that characterizes the noncommutativity, $[\hat{x}^\mu,\hat{x}^\nu]= \Theta^{\mu\nu}$, is related to the background field, and is just an antisymmetric array of c-numbers. However, theories with a c-numbers $\Theta^{\mu\nu}$ suffer from Lorentz-violating effects. Such effects are severely constrained~\cite{MPR}\nocite{Chaichian,HK,CHK,CCL,ABDG,Carlson:2002zb,VN,h2,Bertolami:2003nm}--\cite{Carone:2004wt} by a variety of low energy experiments~\cite{LeExp}. Carone, Zobin, and one of the present authors (CEC)~\cite{CCZ} formulated and studied some phenomenological consequences of a Lorentz-conserving noncommutative QED (NCQED).
The NCQED formulated in~\cite{CCZ} has an underlying noncommutative algebra with $\Theta^{\mu\nu}$ promoted to an operator $\hat{\Theta}^{\mu\nu}$ that transforms like a Lorentz-tensor, and is in the same algebra with $\hat{x}^\mu$. Further studies of NCQED as formulated in~\cite{CCZ} may be found in~\cite{morita}\nocite{CKN}--\cite{Haghighat}.

The aspect of nontrivial commutation relations among spinor coordinates of superspace is also being discussed. Interest in supersymmetric noncommutativity has been stimulated by some recent work (e.g.,~\cite{OoVa}--\cite{Imaanpur}), where it was shown that in Euclidean space  noncommutative supercoordinates could arise from string theory. A number of authors have studied~\cite{Seiberg}--\cite{Klemm} some field theoretical consequences of deformation of $\mathcal{N}=1$ superspace arising from
nonanticommutativity of coordinates $\tta$, while leaving $\btta$'s
anticommuting. This is possible in Euclidean superspace only. In this
talk we present a way to extend the discussion by making both $\tta$ 
and $\btta$ coordinates non-anticommuting in Minkowski superspace. 
The Wess-Zumino Lagrangian $\mathcal{L}_{WZ}$ constructed within our model 
has two extra terms due to nonanticommutativity. There is one extra term
coming from $\int d^2\btta\, \bar{\Phi}*\bar{\Phi}*\bar{\Phi}$ as
compared with Seiberg's result in Euclidean superspace for
$\mathcal{L}_{WZ}$, which is Lorentz invariant. Our result for
$\mathcal{L}_{WZ}$ in Minkowski superspace preserves the Lorentz
invariance and also is manifestly Hermitian~\footnote{See also Ref.~\cite{NC} for a more detailed discussion of this work.}.


\section{The non(anti)commutative algebra}

In constricting our deformed algebra of supersymmetry parameters in Minkowski space~\footnote{We follow conventions of Wess and Bagger~\cite{Wess-Bagger}}, we first require that the deformation be symmetric with respect to chiral and antichiral coordinates. In Minkowski space, we relate $\hat{\bar\tta}^{\dot\al}$ to $\hat{\tta}^\al$ by $\hat{\bar\tta}^{\dot\al}=(\hat{\tta}^\al)^\dagger$. We begin constricting the algebra by first defining the following anticommutator,
\be
\{\hat\tta^\alpha,\hat\tta^\beta\}=C^{\alpha\beta}, \label{com-tta-intro}
\ee
where $C^{\alpha\beta}$ is a symmetric array of c-numbers. Then it also follows that
\be
\{ \hat{\bar\theta}^{\dot \alpha} ,\hat{\bar\theta}^{\dot \beta} \}=
				       \bar C^{\dot\alpha \dot\beta},
\label{com-btta-intro}				        
\ee				 
where $\bar{C}^{\dot\alpha \dot\beta}=(C^{\beta\alpha})^\dagger$. We further make the following simple choice for the yet unconstrained anticommutator of $\hat{\tta}$ and $\hat{\btta}$,
\be
\{ \hat{\bar{\theta}}^{\dot{\alpha}}, \hat\tta^\alpha \}=0.
\label{com-tta-btta-intro}
\ee
We chose the commutators of chiral coordinate 
$\hat{y}^\mu \equiv \hat{x}^\mu+i\hat{\tta}\sigma^\mu\hat{\btta}$, and the antichiral coordinate  
$\hat{\bar{y}}^\mu \equiv \hat{x}^\mu-i\hat{\tta}\sigma^\mu\hat{\btta}$ in such a way that enables us to later write products of chiral fields, and products of antichiral fields, in their standard form. Thus we define
\begin{align} 
[\hat{y}^\mu,\hat{\tta}^\al]&=0, \label{yhat-intro} \\
[\hat{\bar{y}}^\mu,\hat{\btta}^{\dot\al}]& =0\, .  \label{ybarhat-intro}
\end{align}
The choices and results in~(\ref{com-tta-intro})-(\ref{ybarhat-intro}) also constrain the commutation relations of $\hat{y}$ and of $\hat{\bar{y}}$ with themselves. The following condition must be satisfied
\be \label{y-ybar-intro}
[\hat{y}^\mu,\hat{y}^\nu] - [\hat{\bar{y}}^\mu,\hat{\bar{y}}^\nu]
=4(\bcab\hat{\tta}^\al \hat{\tta}^\bt
      - \cab  \hat{\btta}^{\dot\al} \hat{\btta}^{\dot\bt})\smaa \snbb \,.
\ee
Commutators defined in ~(\ref{com-tta-intro})-(\ref{ybarhat-intro}), and the condition~(\ref{y-ybar-intro}) fix the whole algebra of $(\hat{x},\hat{\tta},\hat{\btta})$ coordinates, and we find that
\begin{align} 
\{\hat\tta^\alpha,\hat\tta^\beta\}& =C^{\alpha\beta} \,, &
[\hat{x}^\mu, \hat{\tta}^\al] &= i\cab\smbb\hat{\btta}^{\dot{\bt}}\,, \label{xtta-intro} \\
\{ \hat{\bar\theta}^{\dot \alpha} ,\hat{\bar\theta}^{\dot \beta} \}&= \bcab \,, &
[\hat{x}^\mu, \hat{\btta}^{\dot{\al}}] &= i\bcab\hat{\tta}^\bt \smbb \,, \label{xbtta-intro} \\
\{ \hat{\bar{\theta}}^{\dot{\alpha}}, \hat\tta^\alpha \}&=0 \,, &
[\hat{x}^\mu,\hat{x}^\nu]&=  (\cab \hat{\btta}^{\dot\al} \hat{\btta}^{\dot\bt} - 
                                                            \bcab \hat{\tta}^\bt \hat{\tta}^\al)\smaa\snbb\,. \label{xx-intro} 
\end{align}
Hence, the space-time coordinates $\hat{x}^\mu$ do not commute with each other, or with the spinor coordinates $\hat{\tta}$ and $\hat{\btta}$.

\section{The star product}

We operationally define our theory by finding a suitable star-product. A properly defined star product has to reproduce the underlying noncommutative algebra of deformed 
supersymmetry parameter space in its entirety. We require that the star product satisfy the reality condition, 
\be \label{reality-intro}
(f_1 * f_2)^\dagger = f_2^\dagger * f_1^\dagger
\ee
We will limit the star product to being at most quadratic in deformation parameter $C^{\al\bt}$. This is also the minimum that will allow reproducing the deformed algebra for the supercoordinates. We write down the star product that we use for mapping a product of functions $\hat{f}\hat{g}$ in noncommutative space to a product of functions in commutative space in the following form,
\be
\hat{f}\hat{g} \Rrightarrow f * g = f(1+\mathcal{S})g. \label{S-intro}
\ee
Here $f$ and $g$ can be functions of any of the three sets of variables mentioned above, and the extra operator $\mathcal{S}$ is
\begin{equation} \label{Star-intro}
\begin{split}
\mathcal{S} &=
   		-\frac{C^{\alpha\beta}}{2}
		{\loarrow Q_\alpha}\roarrow{Q_\beta} 
		-\frac{\bar C^{\dot\alpha\dot\beta}}{2}
		\loarrow{\bar{Q}}_{\dot\alpha}\roarrow{\bar{Q}}_{\dot\beta} \\
& \quad  +\frac{\cab \cgd}{8} \loarrow{Q}_\alpha \loarrow{Q}_\gamma   
                                                     \roarrow{Q}_\delta  \roarrow{Q}_\beta 
+ \frac{\bcab \bcgd}{8} \loarrow{\bar{Q}}_{\dot\alpha} \loarrow{\bar{Q}}_{\dot\gamma}
		     \roarrow{\bar{Q}}_{\dot\delta} \roarrow{\bar{Q}}_{\dot\beta} \\
& \quad + \frac{\cab \bcab}{4}\left( \loarrow{\bar{Q}}_{\dot\alpha}\loarrow{Q}_\alpha
			       \roarrow{\bar{Q}}_{\dot\beta} \roarrow{Q}_\beta 
			       -\loarrow{Q}_\alpha \loarrow{\bar{Q}}_{\dot\alpha}
			        \roarrow{Q}_\beta  \roarrow{\bar{Q}}_{\dot\beta} \right)	
\end{split}
\end{equation} 
Here operators $Q$ and $\bar{Q}$ are the supersymmetry generators of canonical supersymmetric theories. The left $\leftarrow$ and right $\rightarrow$ arrows indicate the direction of the action of operators $Q$ and $\bar{Q}$. 
It's easy to verify that the star product presented above indeed reproduces the entire
noncommutative algebra of supersymmetry parameters, and that it satisfies the reality 
condition~(\ref{reality-intro}).
If $f$ and $g$ are functions only of $\tta$, for example, then the star product takes the following simple form, recognizable from~\cite{Seiberg},
\bear
f(\tta)\,*\,g(\tta)&=&
f(\tta){\text exp}\left(-\frac{C^{\al\bt}}{2}\frac{\loarrow\partial}{\partial \ttaa }
	  \frac{\roarrow\partial}{\partial \ttab }\right)g(\tta) \nonumber\\
&=&f(\tta)\left(1-\frac{C^{\al\bt}}{2}\frac{\loarrow\partial}{\partial \ttaa }
		   \frac{\roarrow\partial}{\partial \ttab }
		  -{\text det}C\frac{\loarrow\partial}{\partial \tta\tta }
			      \frac{\roarrow\partial}{\partial \tta\tta }\right)g(\tta) \,,
\label{ft-gt-intro}			      
\eear
where we adopt the following definition:
$
\frac{\partial}{\partial \tta\tta }\equiv \frac{1}{4}\frac{\partial}{\partial\tta_\al}
\frac{\partial}{\partial\ttaa}= \frac{1}{4}\epsilon^{\gamma\eta}
\frac{\partial}{\partial\ttag}\frac{\partial}{\partial \tta^\eta}\,.
$

\section{Supercharges, covariant derivatives, and superfields}

We can now use~(\ref{S-intro}), (\ref{Star-intro}), and the canonical definitions of $Q$ and $\bar{Q}$ to calculate their anticommutators. Thus in noncommutative space we obtain
\begin{align} 
\{Q_\al,Q_\bt \}_*&= -4\bcab \smaa \snbb \frac{\partial^2}{\partial \bar{y}^\mu \partial \bar{y}^\nu} \,,  \label{QQ-intro} \\
\{{\bar Q}_{\dot\al},{\bar Q}_{\dot\bt}\}_*&= -4\cab \smaa \snbb \frac{\partial^2} {\partial y^\mu \partial y^\nu} \,, 
\label{QbarQbar-intro} \\
\{\roarrow{Q}_\al,\roarrow{\bar{Q}}_{\dot\al}\}_* &= 2i\smaa \frac{\partial}{\partial y^\mu} \,. \label{QQbar-intro}
\end{align}
Thus, we see that the first two of the above three anticommutators of supercharges are deformed from their canonical forms. We can still use the canonical definitions for covariant derivatives also, and one can easily verify that their anticommutators are not deformed in noncommutative space defined by~(\ref{xtta-intro})-(\ref{xx-intro}). 

It is important to note that the anticommutators of supercharges and covariant derivatives with each other are not deformed either,
\be
\{ D_\al, Q_\bt\}=\{ \bar{D}_{\dot\al}, Q_\bt \}=\{ D_\al, \bar{Q}_{\dot\bt}\}
=\{\bar{D}_{\dot\al},\bar{Q}_{\dot\bt} \}=0\,. \label{DQ-intro}
\ee
Hence, we can still define supersymmetry covariant 
constraints on superfields as in commutative supersymmetric theory, using the following
defining equations for chiral and antichiral superfields as before,
\bear
{\bar D}_{\dot\al}\Phi (y,\tta) &=& 0\,,\label{chiral-intro}\\ 
D_\al {\bar \Phi}({\bar y},\btta)&=&0\,.\label{antichiral-intro}
\eear
On the other hand, from~(\ref{QQ-intro})-(\ref{QQbar-intro}) it is also clear that the star product is not invariant under $Q$ or $\bar{Q}$~\cite{Seiberg,Ferrara}. Therefore, the star product breaks whole of the supersymmetry, and neither $Q$, nor $\bar{Q}$ are symmetries of noncommutative space described by~(\ref{xtta-intro})-(\ref{xx-intro}).

\section{The Wess-Zumino Lagrangian}

Chiral $\Phi(\hat{y},\hat{\tta})$, and antichiral $\bar{\Phi}(\hat{\bar{y}},\hat{\bar{\tta}})$ fields as defined by~(\ref{chiral-intro}) and~(\ref{antichiral-intro}) can be expanded as a power series in $\hat\tta$ and $\hat\btta$. The series will still have terms with no more than two powers of $\hat\tta$ and $\hat\btta$, 
\bear
\Phi(\hat{y},\hat{\tta}) &=& A(\hat{y}) + \sqrt{2}\hat{\tta}\psi(\hat{\tta}) +
			     \hat{\tta}\hat{\tta}F(\hat{y})\,,\label{chiral_s-intro}\\
\bar{\Phi}(\hat{\bar{y}},\hat{\bar{\tta}}) &=& A(\hat{\bar{y}}) + 
				 \sqrt{2}\hat{\btta}\bar{\psi}(\hat{\bar{y}})+
				 \hat{\btta}\hat{\btta}\bar{F}(\hat{\bar{y}})\,.\label{antichiral_s-intro}
\eear 
Then one can use the star product defined by~(\ref{S-intro}), (\ref{Star-intro}) to calculate products of chiral and antichiral fields. For a product of two chiral fields we obtain
\begin{equation} \label{chiral-1-2-intro}
\begin{split}
\Phi_1(y,\tta)\, * \,\Phi_2(y,\tta)&={\Phi_1}(y,\tta)\Phi_2(y,\tta)
-C^{\al\bt}\psi_{1\al}\psi_{2\bt} -{\textrm{det}}C F_1F_2 \\
& \quad + \sqrt{2}\tta^\gamma\cab\big[\epsilon_{\bt\gamma}(\psi_{1\al}F_2 - \psi_{2\al}F_1)  \\
& \quad   +\bcab \smaa\sigma_{\gamma\dot\bt}^\nu(\partial_\mu A_1 \partial_\nu \psi_{2\bt} 
					 - \partial_\mu A_2 \partial_\nu \psi_{1\bt} ) \big]\\
& \quad +\tta\tta\big[ 2\bar{C}^{\mu\nu}\partial_\mu A_1 \partial_\nu A_2 \\
& \qquad\quad \, +\cab\bcab \smaa\sigma_{\bt\dot\bt}^\nu(\partial_\mu A_1 \partial_\nu F_2 
					 - \partial_\mu A_2 \partial_\nu F_1 )\big] \,,
\end{split}
\end{equation}
with $\partial_\mu$ defined as $\partial / \partial y^\mu$. We can see right away that the right hand side of~(\ref{chiral-1-2-intro}) is a chiral field. Thus the star product maintains the chirality of products of chiral fields, and it can be checked that it also maintains the antichirality of products of antichiral fields. One can also check explicitly that the reality condition is indeed satisfied: $\overline{(\Phi_1 * \Phi_2)} = \bar{\Phi}_2 * \bar{\Phi}_1$.

We must note that the star product~(\ref{Star-intro}) used in our studies is not associative. However, this interesting feature causes little trouble after making a natural modification of the Weyl ordering procedure, generalizing it for noncommutative, nonassociative products, 

\begin{equation} \label{Weyl-1}
\begin{split}
{\textrm{\bf W}}(f_1 (f_2 f_3)) & \equiv \frac{1}{6}\big[ f_1(f_2f_3)+f_2(f_1f_3) + f_2(f_3f_1)+
				                     f_1(f_3f_2)+f_3(f_1f_2) + f_3(f_2f_1) \big]\\
&= \frac{1}{6}\big[ f_1(f_2f_3+f_3f_2) + f_2(f_1f_3+f_3f_1) + f_3(f_1f_2 + f_2f_1) \big] \,.
\end{split}
\end{equation}
and similarly  for ${\textrm{\bf W}}((f_1 f_2) f_3)$.
One can follow this by Weyl ordering the result in the normal way and find that
\be \label{WW}
{\textrm{\bf W}} \left[{\textrm{\bf W}} (f_1 (f_2 f_3))\right]=
{\textrm{\bf W}} \left[{\textrm{\bf W}} ((f_1 f_2) f_3)\right] 
\equiv  {\textrm{\bf w}}  (f_1 f_2 f_3) \,.
\ee


\noindent Thus we limit our discussion to double Weyl ordered products of fields, and we write down the Wess-Zumino Lagrangian with double Weyl ordered terms. We would like to note that a similar procedure was introduced by Seiberg in Ref.~\cite{Seiberg} to deal with the fact that the star product used in his model was noncommutative. Thus, in Ref.~\cite{Seiberg} the discussion was limited to products of fields that were Weyl ordered. 

We find the following simple result for the Wess-Zumino Lagrangian with one chiral $\Phi$,
and one  antichiral field $\bar\Phi$, 
\begin{equation} \label{W-Z-intro}
\begin{split}
\mathcal{L}_{WZ}&={\textrm{\bf w}}\Big[\int d^2\tta\tta d^2 \btta\btta \,\bar{\Phi}*\Phi
+\int d^2\tta\,\left(\frac{1}{2}m\Phi * \Phi +\frac{1}{3}g\Phi *\Phi * \Phi\right) \\
&\quad\quad\quad\quad\quad\qquad \qquad\quad\;\>+\int d^2\btta\,\left(\frac{1}{2}m\bar{\Phi} * \bar{\Phi} +\frac{1}{3}g\bar{\Phi} * \bar{\Phi} * \bar{\Phi}\right) \Big] \\
&=\mathcal{L}(C=0)-\frac{1}{3}g{\textrm{det}}C F^3 -\frac{1}{3}g{\textrm{det}}\bar{C}\bar{F}^3 
+\textrm{total derivatives}\,.
\end{split}
\end{equation}
Here {\bf w[ ]} means double Weyl ordering, $\mathcal{L}_{WZ}(C=0)$ is the term representing the canonical part of the Lagrangian, and $F$, $\bar{F}$ are the F-terms in chiral, and antichiral superfields respectively. The total derivatives indicated in~(\ref{W-Z-intro}) arise due to coordinate transformation form $y$, and $\bar{y}$ to $x$, and will cancel in the action.

\section{Summary}

We have studied the consequences of deformation of $\mathcal{N}=1$ Minkowski superspace arising from nonanticommutativity of coordinates $\tta$, and $\btta$. We presented a consistent algebra for the supercoordinates, and found a star-product that reproduces all the coordinate commutation relations. We used this star product to define multiplication of arbitrary functions. The star product developed in our studies is real, meaning it maintains the standard relations obeyed by involutions of products of functions. As a consequence, the star product preserves the hermiticity of a Weyl ordered product of functions. Any Lagrangian extended to noncommutative space using star-products and Weyl ordering will necessarily remain Hermitian. 
Further, the star-product maintains the chirality properties of products of both chiral, and antichiral fields. We also made a natural generalization of the Weyl ordering procedure, to take into account the interesting feature of nonassociativity of the star product. 


The Wess-Zumino Lagrangian obtained in our model is Hermitian, and gains only Lorentz invariant correction due to non(anti)commutativity. We note that only corrections up to second order in deformation parameter $C$ are presented in~(\ref{W-Z-intro}). Higher order corrections due to noncommutativity may very well destroy the nice feature of Lorentz invariance, although the Lagrangian will remain Hermitian. 

We will study the problem of formulating a gauge field theory with an underlying algebra~(\ref{xtta-intro})-(\ref{xx-intro}) in a future work. Another interesting problem is to study if the deformation of SUSY coordinate algebra considered in this work can arise from super-string theory.
 
\section*{Acknowledgments}
We thank Chris Carone and Marc Sher for useful discussions. We also thank the NSF for support under Grant PHY-0245056.

\end{document}